\documentclass[twocolumn,arguments,twocolappendix]{aastex631}

\usepackage{amsmath}
\let\splitbox\relax
\usepackage[export]{adjustbox}[2011/08/13]
\usepackage{hyperref}
\defcitealias{Rogers2017}{RM17}
\received{July 29, 2022}
\revised{October 5, 2022}
\accepted{November 5, 2022}

\submitjournal{ApJ}

\shorttitle{AASTeX v6.3.1 Sample article}
\shortauthors{Varghese et al.}
\graphicspath{{./}{figures/}}

\begin{document}

\title{Chemical Mixing Induced by Internal Gravity Waves in Intermediate Mass Stars}

\author{A. Varghese}
\affiliation{Newcastle University \\
Newcastle Upon Tyne, United Kingdom}


\author{R. P. Ratnasingam}
\affiliation{Nicolaus Copernicus Astronomical Center \\
Polish Academy of Sciences, Poland}

\author{R. Vanon}
\affiliation{Newcastle University \\
Newcastle Upon Tyne, United Kingdom}

\author{P. V. F. Edelmann}
\affiliation{Los Alamos National Laboratory \\
Los Alamos, United States}

\author{T. M. Rogers}
\affiliation{Newcastle University \\
Newcastle Upon Tyne, United Kingdom}
\affiliation{Planetary Science Institute, Tucson, AZ, USA}




\begin{abstract}

Internal gravity waves (IGWs) can cause mixing in the radiative interiors
of stars. We study this mixing by introducing tracer particles into two-
dimensional (2D) hydrodynamic simulations. Following the work of \cite{Rogers2017}, we extend our study to different masses (3 M$_{\odot}$, 7 M$_{\odot}$ and 20 M$_{\odot}$) and
ages (ZAMS, midMS and TAMS). The diffusion profiles of these models are influenced by
various parameters such as the Brunt-Väisälä frequency, density, thermal damping, the geometric effect and the frequencies of waves contributing to these mixing profiles. We find that the mixing profile changes dramatically across age. In younger stars, we noted that the diffusion coefficient increases towards the surface, whereas in older stars the initial increase in the diffusion profile is followed by a decreasing trend. We also find that mixing is stronger in more massive stars. Hence, future stellar evolution models should include this variation.  In order to aid the inclusion of this mixing in one-dimensional (1D) stellar evolution models, we determine the dominant waves contributing to these mixing profiles and present a prescription that can be included
in 1D models.

\end{abstract}



\section{Introduction} \label{sec:intro}

Chemical mixing plays a major role in stellar evolution. It can transport the chemical elements generated in the nuclear burning region to the surface, creating observed abundance variations.  In massive stars, it influences the lifetime of the star by bringing fresh fuel into the burning region leading to a more massive Helium core.
Standard stellar evolution theory considers few transport mechanisms or instabilities in the stable radiative zone and yet, very successful in explaining many of the observed properties of stars, such as the mass-luminosity relation, the existence of the main sequence and the evolution towards the giant branch \citep{zahn94}.

However, a series of observed anomalies suggested the need to consider additional mixing processes near the convective-radiative boundary and in the radiation zone; some amongst them are the N/C and $^{13}$C/$^{12}$C enrichment in low mass stars on the upper red giant branch \citep{char1994}, N abundances in B stars \citep{gies}, He abundances in O and B stars near the main sequence \citep{lyu_75, lyu_77, lyu_89}, He discrepancy in O type stars \citep{Herr}, He and N enrichments in OB supergiants \citep{wal_1976}, Li abundances in population II low mass giants \citep{pil_1993}, and  C, N, O abundances in globular cluster giants \citep{kraft}. Some of the additional physical processes that plausibly cause mixing in these regions are convective overshoot at the convective - radiative boundary \citep{zan1991, herwig2000}, rotation \citep{zahn_92, MZ_98, Mm2000, Heger2T}, shear induced turbulence \citep{zahn_92, T&Z_97,mad_2003,mathis_2004,mathis_2018,KG, park_2020} internal gravity waves (IGWs) (\citeauthor{Press} \citeyear{Press}; \citeauthor{gs1991} \citeyear{gs1991}; \citeauthor{CT_2007} \citeyear{CT_2007}; \citeauthor{ Rogers2017} \citeyear{Rogers2017}, hereafter \citetalias{Rogers2017}) and magnetic fields \citep{m&m_2005, heger2005}.
  Over the years stellar evolution models focused on the inclusion of these additional mechanisms and  using the 
  different  prescriptions given by \cite{kip_1970},  \cite{es_1976}, \cite{zahn_92},  \cite{MZ_98}, \cite{herwig2000}, \cite{spirit_2002}, etc. these mechanisms are implemented in stellar evolution models such as MESA, FRANEC, STERN, GENEC, STAREVOL, etc \citep{mesa1, mesa2, mesa3, mesa4, mesa5,  starevol1, geneva, brott, franec_2013}. These models were successful in explaining  many observed anomalies such as the observed Nitrogen enhancement in the early universe \citep{hirshi_2007}, surface enrichment of He and C in Main sequence stars \citep{hl_2000}, blue side of the Li dip and low Li abundances in subgiants  \citep{pal_2003}.
  
  Mixing by convective overshooting is one of the oldest problems studied in stellar astrophysics. Overshooting occurs when the motion of  matter extends beyond the boundary determined by either the Ledoux or Schwarzschild  stability criteria and continues into the stable layer. Convective matter overshoots outward from the convective core to the radiative envelope in stars more massive than the sun, while for solar-type stars ($0.85 \leq M/M_{\odot} \leq 1.2$) the process is inward, from the convective envelope, leading it to also be called "undershooting"; in either case, the phenomenon can play an important role in stellar evolution. 
Studies by \cite{alongi} showed that convective undershooting modifies the chemical abundance and also explains the bump in the luminosity function of RGB stars in globular clusters. Convective core overshooting extends the lifetime of core H burning by increasing the availability of fuel within the core, lead to a more massive He core and enhancing the luminosity in the later evolutionary track \citep{m_1976, klaus}. 



With the recent advancements in asteroseismology, it was found that the distinct signatures in the period spacings of gravity mode spectra can provide information on the extent to which chemical elements are mixed and the nature of mixing in the transition layer between a convective and radiative zone and in the radiative interiors \citep{sies, noel, CD, may, seb}. 
Asterosesmic studies also reveal a more massive core than expected by the standard stellar evolution theory suggesting the need of extra mixing in the earlier evolutionary stages \citep{charpet, giam}. Callibration of mixing profiles based on asteroseismic modelling provides an important input for the improvement of stellar evolution models \citep{may_nature}. 

Rotation strongly influences stellar evolution  and is considered to be a major ingredient in stellar models and may account for the observed abundance anomalies in massive stars.
\cite{edd} proposed that meridional circulation
can induce mixing in stellar interiors and later studies  have indicated that this circulation could explain the C to N ratio in many early type stars \citep{pas}.  Internal mixing by rotation could explain the Li and Be abundances in low mass stars \citep{pin1989}, N/C enhancement seen in OB stars \citep{lyu}, the overabundance of He in O stars \citep{Herr}, Li abundances in evolved stars \citep{mart_2011} and the surface enrichment of O7-8 giants \citep{martin}. More mixing is contributed by differential rotation that generates turbulent motions in stellar interiors, which give rise to various rotationally induced instabilities like  Goldreich-Schubert-Fricke Instability (\citeauthor{gs} \citeyear{gs}), ABCD-instability \citep{spirit} and shear instabilities \citep{zahn_92, brug, PL, garaud2020,CG2021, park2021, prat_mathis}. These instabilities play a major role in mixing the CNO materials close to the surface in subgiant and giant branch models \citep{pin1989}.  
Although models that included rotation could explain many of the observed anomalies, they failed to account for some of the observed features like the abundance anomalies in red giants \citep{palacios}, the flat solar rotational profile measured by helioseismology \citep{brown, M_Z_1997} and the nitrogen abundance in massive stars \citep{brott, A2014}. Models including rotational induced transport mechanisms such as the meridional circulation and shear instabilities predicted a rapidly rotating stellar core in sub-giants \citep{ceil}, red giants \citep{Egg_2012,marq, matteo_2014} and $\gamma$ Doradus main-sequence intermediate-mass stars \citep{ozz_2019} contradicting the core rotation revealed by asteroseismic observations. This also suggests the need to consider additional transport mechanisms in the stellar interiors.


Apart from rotation, magnetism could also be a source of extra mixing in stars. \cite{m&m_2005} studied a magnetic model with the inclusion of meridional circulation and found a significant enrichment in N and He compared to rotating models without an internal magnetic field. Later studies argued that it is likely for magnetic buoyancy to transport materials in AGB and RGB stars \citep{busso, nor, diego}.

Recently thermohaline mixing,  which occurs because of the molecular weight inversion, was proposed to influence the surface abundances in low mass red giants (\citeauthor{cz2007} \citeyear{cz2007}; \citeauthor{egg} \citeyear{egg}). It could explain the $^3$He, Li, C and N abundance in the low mass stars above the RGB bump and the high Li abundances in low mass AGB stars. \citep{char_l, stan, lad_2011}. \cite{CL} investigated the role of this mixing along with rotational mixing and internal gravity fields in low-mass stars. They confirmed that this mixing could explain the decrease in He abundance in RGB stars with an initial mass less than 1.5M$_{\odot}$. 

Internal Gravity Waves (IGWs) can also cause extra mixing in the radiative  interiors \citep{Press, mon94, ms96, ms93, Rogers2017}. IGWs are naturally occurring waves that propagate in stably stratified fluids with gravity as the restoring force. In stars, IGWs are generated at the convective - radiative interface either by bulk excitation through Reynolds stress \citep{Kumar,belkam_2009,samadi_2010, LQ} or by direct excitation through plumes \citep{hurl,mons_2000,pincon2016}.  
\citet{Press} considered that the turbulent motion in the convective zone is characterised by eddies and studied the turbulent mixing by IGWs in stars with a convective envelope. He proposed that IGWs could explain the solar neutrino discrepancy. \cite{gs1991} argued that the turbulence generated by the internal gravity  waves  propagating in radiative interiors can explain the Li gap in F type stars. \cite{mon94} used plume morphology based on convective penetration (\citeauthor{zan1991} \citeyear{zan1991}) to describe the perturbation at the boundary and explained the observed  Li abundance in the Sun (\citeauthor{ms96} \citeyear{ms96}). \citetalias{Rogers2017} studied the mixing by IGWs in massive stars and determined that it can be treated as a diffusive process with the diffusion profile set by the square of the wave amplitude.
Theoretical models including rotation and IGWs were successful in providing insight into the surface Li abundance in low mass stars \citep{TC5, CT_2007}.

In this paper, we focus on the mixing by IGWs in stars with a convective core and a radiative envelope. We extend the studies of \citetalias{Rogers2017} to obtain the mixing profiles for stars of higher masses and different ages. We achieve this by running 2D simulations using a background reference model obtained from Modules for Experiments in Stellar Astrophysics (MESA) \citep{mesa1, mesa2, mesa3, mesa4, mesa5}. We then determine the dominant wave contributing to these mixing profiles using the prescription given by \citetalias{Rogers2017}. Section \ref{sec:style} explains the 2D simulations and the tracer particle simulations done to obtain the mixing profiles. Section \ref{sec:floats} introduces our results on the mixing profiles for the different models including a brief description of the theoretical calculations used in this work and Section \ref{conclsn} summarises our findings.

\section{Background and Numerical Methods} \label{sec:style}

\subsection{Two-Dimensional Simulation} \label{2d}
We generated the stellar models for three masses (3 M$_{\odot}$, 7 M$_{\odot}$, 20 M$_{\odot}$) at three ages (Zero Age Main Sequence (ZAMS, core hydrogen mass fraction, X$_{c}$ = 0.7), mid-Main Sequence (midMS, X$_{c}$ = 0.35), Terminal Age Main Sequence (TAMS, X$_{c}$=0.01) from MESA. We set the overshooting parameter to 1.8 and the convective overshoot profile to exponential, implemented the mass loss through Van Loon stellar wind scheme and considered zero rotation in our models. The inlists used to generate the models are given in Appendix \ref{app} and also available in \href{https://zenodo.org/record/2596370\#.Yn5quDnMJUR}{zenodo}. 

\begin{figure*}
\includegraphics[width=1\textwidth]{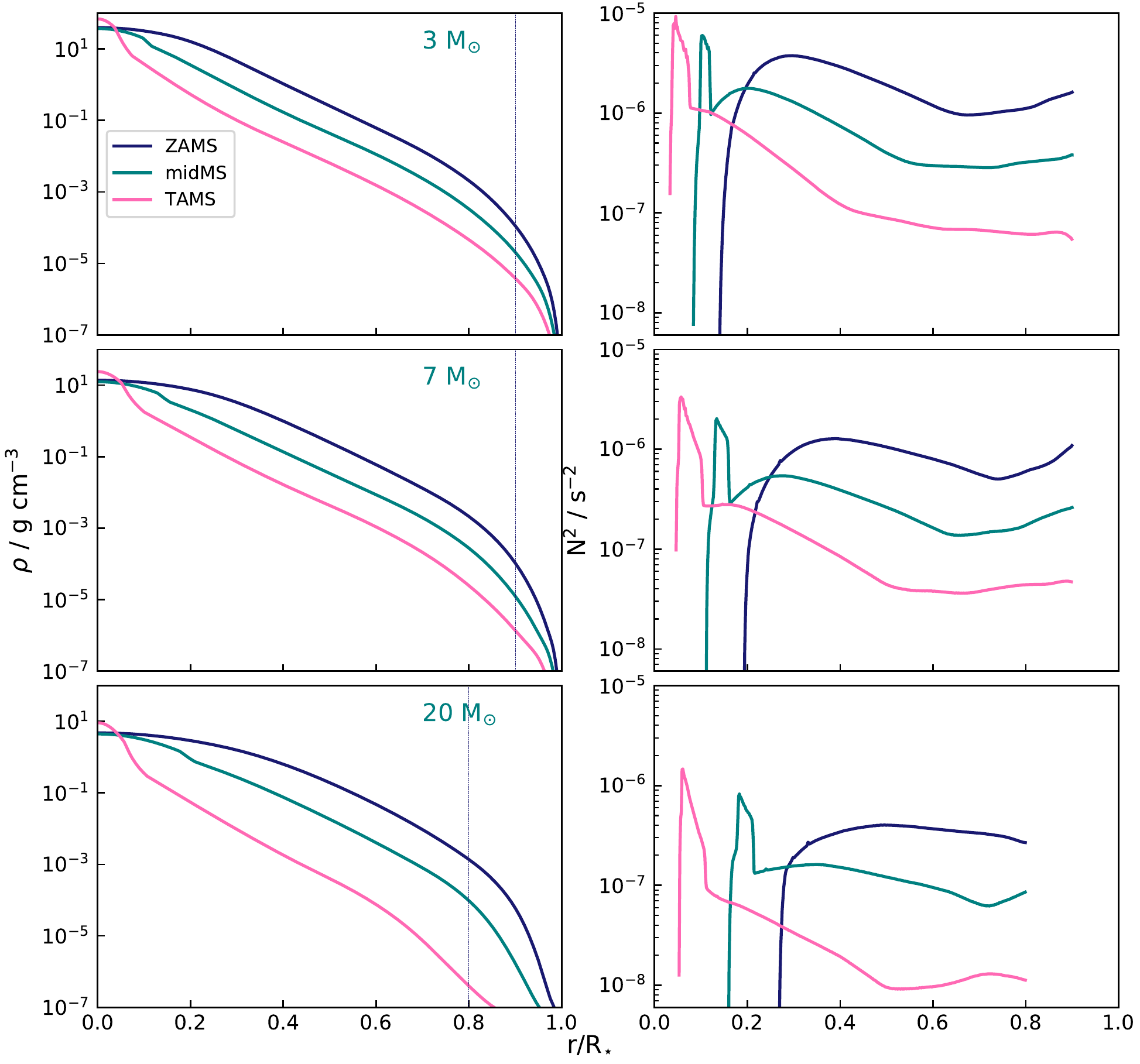}
\caption{ Density (left panel) and Brunt-V\"{a}is\"{a}l\"{a} frequency (right panel) and  as a function of fractional radius for 3 M$_{\odot}$ (top panel), 7 M$_{\odot}$ (middle panel), 20  M$_{\odot}$ (bottom panel) at ZAMS, midMS and TAMS. The vertical dashed line in the density plot indicates the cut off radii of the simulation domain.}
\label{den}
\end{figure*}

We calculated the Brunt-V\"{a}is\"{a}l\"{a} frequency, N based on the work of \cite{R2013},

\begin{equation}\label{bvf_r}
    N^{2}=\frac{\bar{g}}{\bar{T}}\left(\frac{d\bar{T}}{d\bar{r}} - (\gamma - 1)\bar{T}h_{\rho}\right).
\end{equation}
where $\bar{g}$, $\bar{T}$, $\gamma$  and $h_{\rho}$ are the reference state gravity, temperature, the adiabatic index and the negative inverse density scale height. The location of the convective-radiative interface is determined by the sign of the Brunt-V\"{a}is\"{a}l\"{a} frequency where
 \begin{equation}
     N^{2}>0
 \end{equation}
 is the radiation zone and
  \begin{equation}
     N^{2}<0
 \end{equation}
 is the convection zone.
 The background density and the Brunt-V\"{a}is\"{a}l\"{a} frequency profiles of all the models considered in this work are shown in Fig. \ref{den}.
\begin{figure*}
\includegraphics[width=1\textwidth]{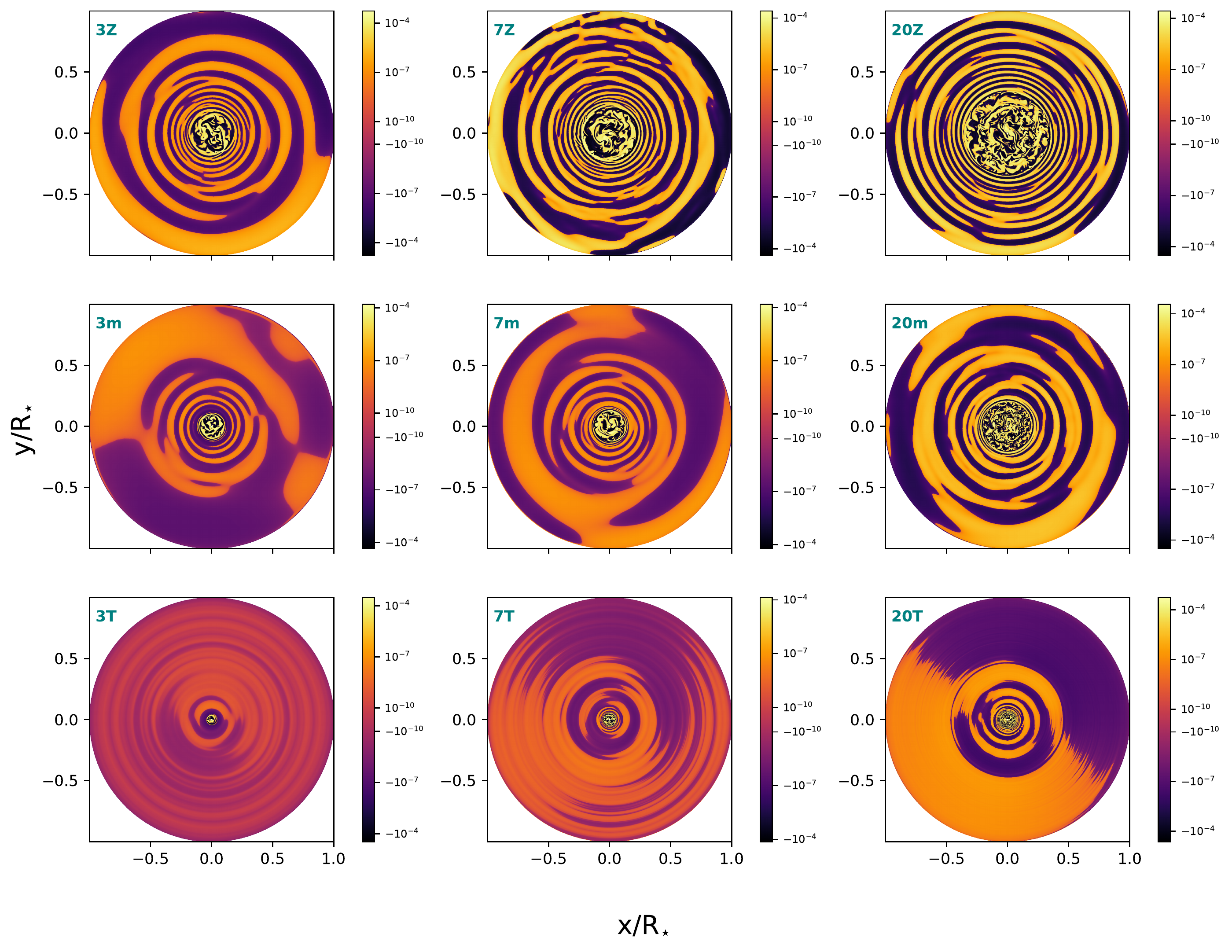}
\caption{Vorticity  for 3 M$_{\odot}$ (left), 7 M$_{\odot}$ (middle) and 20 M$_{\odot}$ (right) at ZAMS (top panel), mid-MS (middle panel) and TAMS (bottom panel. }
\label{vort_37}
\end{figure*}

 
 To study IGWs in stellar interiors, we solved the Navier-Stokes equations in the anelastic approximations using 2D hydrodynamic simulations considering an equatorial slice of the star with stress-free, isothermal and impermeable boundary conditions \citep{R2013}. The numerical model followed in this work is similar to that of \cite{R2013} and the 2 D simulations are run by R.P Ratnasingam (more details on the simulations can be found in \cite{phdthesis}, Ratnasingam et al. 2022 in prep).

Fig. \ref{vort_37} shows the time snapshots of vorticity  for all the models considered in this work.
 
The thermal diffusivity ($\kappa$) and viscosity ($\nu$) are set to constant values shown in Table. \ref{table_1}. Our objective was to attain the highest  Reynolds number, Re,
 \begin{equation}
    \mathrm{Re} =  \frac{v_{rms}L}{\nu}.
 \end{equation}
for each model studied. Here $v_{rms}$ is the root mean squared velocity averaged over the convection zone and $L$ is the characteristic length scale (the radial extent of the convection zone in this case). We also aimed to attain convective velocity ratios for different masses from our simulations which are comparable to those predicted by the mixing-length theory (MLT) velocities in MESA for any given age. (We chose to compare across masses as the uncertainties in the MLT velocities in MESA are larger across ages). We therefore chose the values such that the above are satisfied whilst maintaining numerical stability.
 The simulation is run for an average time of approximately, $2.7\times 10^{7}$s for the models studied. 
 \begin{table}[ht]
  \centering

\begin{tabular}{ccc} 
 \hline
 Model & $\kappa$, $\nu$ / $10^{12}$ $\mathrm{cm^2 s^{-1}}$ & Convective turnover \\ [0.5ex] 
 & & times \\
 \hline
3M$_{\odot}$ ZAMS&  5 & 123 \\
3M$_{\odot}$ midMS & 5 & 91\\
3M$_{\odot}$ TAMS & 5 & 105 \\
7M$_{\odot}$ ZAMS & 5 &62\\
7M$_{\odot}$ midMS & 5 &70 \\
7M$_{\odot}$ TAMS & 2.5 &67\\
20M$_{\odot}$ ZAMS & 8 &40 \\
20M$_{\odot}$ midMS &8 &58 \\
20M$_{\odot}$ TAMS & 5 &52\\ 
 \hline
\end{tabular}
\caption{Thermal and viscous diffusivities of the different models used in our simulation along with the total time interval in terms of convective turnover times considered for our analysis.}
\label{table_1}
\end{table}





The simulation domain extends up to $90 \, \%$ of the total stellar radius for all the 3 M$_{\odot}$ and 7 M$_{\odot}$ models and  up to $80 \, \%$ of the total stellar radius  for all the 20 M$_{\odot}$ models. The cut-off radius is determined such that the density does not vary more than $\sim$ 6 orders of magnitude as we move from the inner to the outer boundary to maintain numerical stability. As evident from Fig. \ref{den}, this is satisfied only until $0.9R_{\star}$ for 3 M$_{\odot}$ and 7 M$_{\odot}$ models  and up to $0.8R_{\star}$ for 20 M$_{\odot}$ models. We therefore chose these respective domains for our simulations. For our analysis, we chose the velocity data from a time,  $4\times 10^{6}$s for all the 3 M$_{\odot}$ and 7 M$_{\odot}$ models and from $6\times 10^{6}$s\footnote{We found that the IGWs show a steady state evolution from this value in our 2D hydrodynamic simulation. Steady- state evolution refers to a state when the amplitudes of the IGWs remain constant over time.} for all the 20 M$_{\odot}$ models. The total time window considered is given in terms of convective turnover times in Table. \ref{table_1}.  
We then used this velocity data which is saved at a regular time interval to study the mixing by IGWs in stellar interiors.

\subsection{Tracer particle simulation}
To measure the mixing by IGWs in stellar interiors, we introduce $\mathcal{N}$ tracer particles into our simulation and track them over a time $T$. This procedure is carried out in post-processing  and we use the relations given by \citetalias{Rogers2017} to calculate the diffusion coefficient $D(r,\tau)$ at radius $r$, for a given time difference, $\tau$.
 \begin{align} \label{dif}
 D(r,\tau) = \frac{Q(r,\tau)}{2\tau n(r,\tau)}-\frac{P(r,\tau)^{2}}{2\tau n(r,\tau)^{2}}.
 \end{align} 
 where $n(r,\tau)$, is the number of sub-trajectories starting at $r$ with a duration of $\tau$, $P(r,\tau)$ is the sum of the lengths of the sub-trajectories and $Q(r,\tau)$ is the sum of the square of these sub-trajectories. More details on the calculation of $n$, $P$, $Q$ and $D(r,\tau)$ can be found in \citetalias{Rogers2017}. 

\begin{figure}
\includegraphics[width=0.45\textwidth]{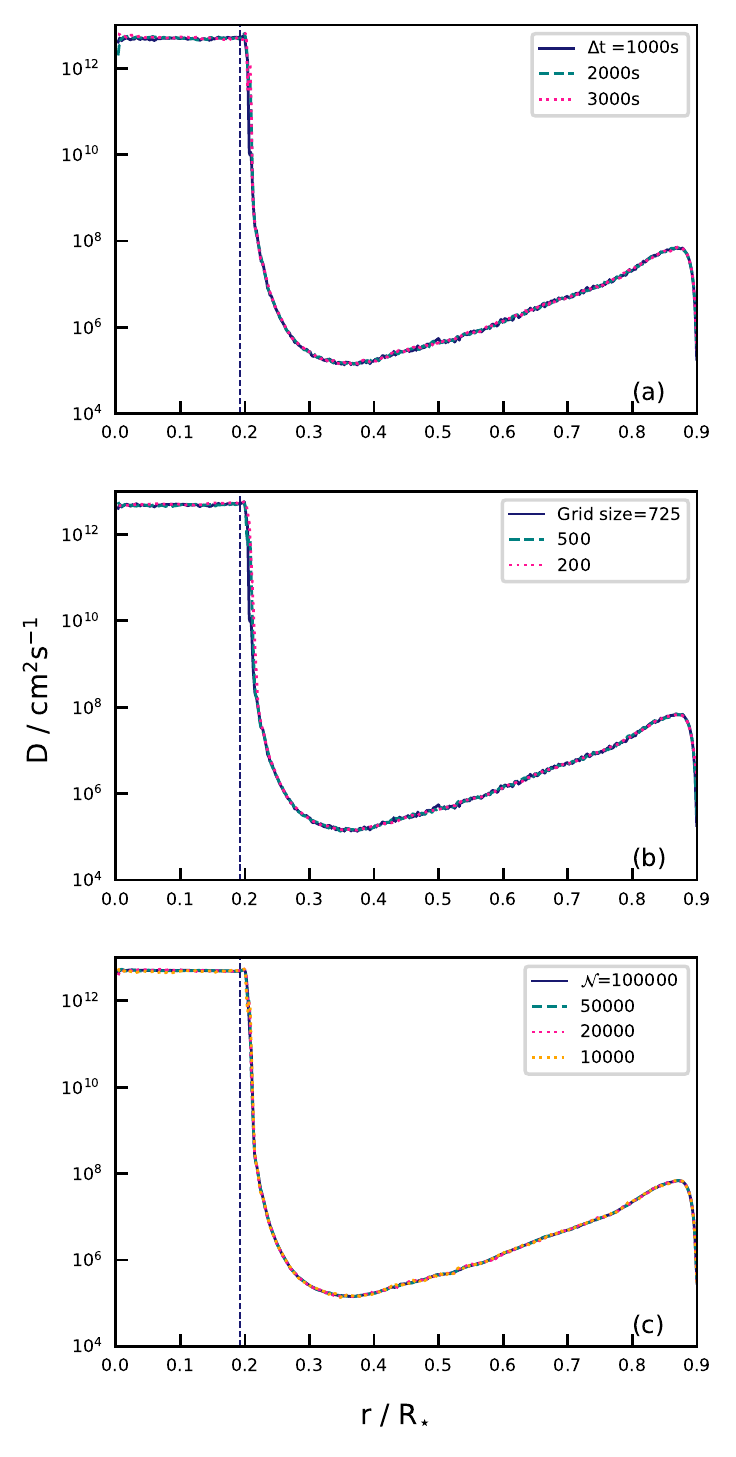}
\caption{Radial diffusion profiles for varying parameters (a) Velocity time step ($\Delta$t) (b) Radial grid size (c) Number of particles ($\mathcal{N}$)  for 7 M$_{\odot}$ ZAMS model at $\tau$ $=$ $1.2\times 10^{6}$s. The blue vertical dashed line indicates the convective - radiative interface}
\label{resln}
\end{figure}


We then plotted the diffusion coefficients calculated using the above equation as a function of radius. We varied different parameters such as the number of particles, radial grid size and velocity timesteps in our simulation to check for numerical convergence and obtained the diffusion coefficient for each set of parameters. 
Fig. \ref{resln} shows the radial diffusion profiles with (a) different velocity time steps $\Delta$t (b) different radial grid sizes and (c) different number of particles $\mathcal{N}$ ,  for a 7 M$_{\odot}$ ZAMS model. In each case, the profiles appear converged with respect to the variables changed. We extended the same treatment to all the other stellar models and found the radial diffusion profile to be already converged for a time resolution of $1000$s, radial resolution of $500$ and particle resolution of 10000. This is consistent with \citetalias{Rogers2017} who found the mixing profile robust to parameter changes. Henceforth, we choose these values for all our further analysis and obtain the radial diffusion profile at a time difference \footnote{ The smaller $\tau$ has contributions from more time steps compared to a larger value of $\tau$. As an example, consider T$= 100$ with time step $= 1$. Then, for $\tau =5$, it has contributions from $5-0$, $6-1$, $7-2$ and so on, whereas $\tau=97$ can result only from $100-3$, $99-2$ and $98-1$.} of $1.3\times 10^{7}$s for  all the three masses. 

\section{Results} \label{sec:floats}

We found the general trend of the diffusion profile to be increasing from the convective - radiative interface towards the surface as seen in Fig. \ref{resln} consistent with the results of \citetalias{Rogers2017}. However as the star ages, its properties such as the density and the Brunt-V\"{a}is\"{a}l\"{a} frequency change is evident from Fig. \ref{den}. These changes also affect the diffusion coefficient. Here, we investigated the dependence of age and mass on the radial diffusion profiles. 

\subsection{Age dependencies}\label{age}
\begin{figure}
\includegraphics[width=0.45\textwidth]{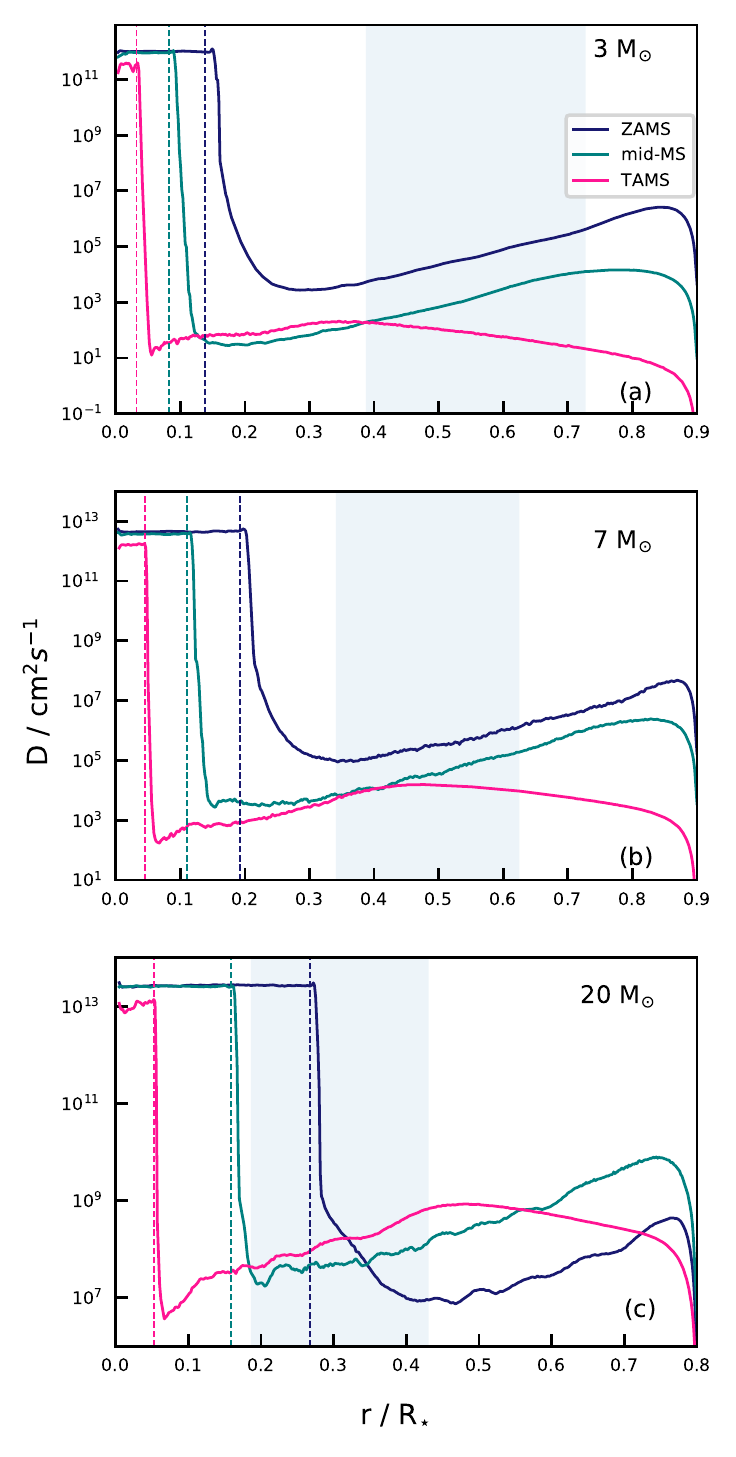}
\caption{Diffusion coefficient as a function of fractional radius at ZAMS, mid-MS and TAMS for (a) 3 M$_{\odot}$, (b) 7 M$_{\odot}$ and (c) 20 M$_{\odot}$ models. The vertical dashed line represents the convective-radiative interface for each models. The shaded region indicates the location of turning points of the TAMS model for the range of frequencies $4 - 14 \mu \mathrm{Hz}$.}
\label{l_3ZAMS}
\end{figure}

Fig. \ref{l_3ZAMS} shows the radial diffusion profiles at ZAMS (Blue), midMS (green) and TAMS (pink) for 3 M$_{\odot}$, 7 M$_{\odot}$ and 20 M$_{\odot}$  models.
In general, we find that the mixing profile varies significantly across the ages for all the masses studied. The amplitude of the diffusion profile depends on several factors such as the wave driving, density stratification, thermal diffusivity, the Brunt-V\"{a}is\"{a}l\"{a} frequency and the geometric effect. Among these, density stratification and thermal diffusivity dominate in both ZAMS and midMS stars, hence the diffusion coefficient increases with stellar radius following the density stratification modulated by the thermal diffusion. 

However, the mixing profile at TAMS shows an increasing trend from the convective - radiative interface followed by a $\it{decreasing}$ trend towards the surface which is considerably different from that of the younger stars. To understand this behaviour, we looked at the following second - order differential equation for wave propagation, 
\begin{align}\label{red_LT}
0  = & \frac{\partial^2 \alpha}{\partial r^2} + \left[\frac{N^2}{\omega^2} - 1 \right]\frac{m^2}{r^2}\alpha +\frac{1}{2}\left[\frac{\partial h_{\rho}}{\partial r} - h_{\rho} ^{2}  +\frac{h_{\rho}}{r}\right]\alpha  \nonumber  \\
 & + \frac{1}{4r^2}\alpha
\end{align}
 which is obtained by reducing the linearised 2D hydrodynamic equations considering no thermal or viscous diffusion with $\alpha = v_{r}\bar{\rho}^{\frac{1}{2}}r^{\frac{3}{2}}$ and m is the 2D Fourier basis wavenumber. In younger stars, the first two terms dominate over most of their radii. However, in older stars, the density term,
 \begin{equation}
    \frac{1}{2}\left[\frac{\partial h_{\rho}}{\partial r} - h_{\rho} ^{2}  +\frac{h_{\rho}}{r}\right] ,
\end{equation}
 becomes dominant at larger radii.
 
 We define turning point as the radius at which the ratio of the oscillatory term,
\begin{equation}
    \left[\frac{N^{2}}{\omega^{2}} -1\right] \frac{m^{2}}{r^{2}}.
\end{equation}
 to that of the density term $\left(\frac{1}{2}\left[\frac{\partial h_{\rho}}{\partial r} - h_{\rho} ^{2}  +\frac{h_{\rho}}{r}\right] \right)$ is equal to 1 \citep{Rathish2020} and it is located at a lower fraction of the total radius in older stars. Waves lose their  wave like behaviour when this ratio is less than one and as a result they experience extra damping.
This  leads to waves becoming evanescent towards the surface and leads to the diffusion profile we see in Fig. \ref{l_3ZAMS}. The shaded region in the plot indicates the location of turning points for the TAMS models using the range of frequencies ($4 - 14 \mu \mathrm{Hz}$) determined in Section \ref{dom_freq}.    

Another factor contributing  to the diffusion profiles of TAMS models is the Brunt-V\"{a}is\"{a}l\"{a} frequency spike near the convective-radiative interface left behind by a steep composition gradient as the star evolves from ZAMS (Fig. \ref{bvf_r}).
This peak damps many waves near the convective - radiative interface and traps the higher frequencies resulting in fewer waves contributing to mixing, thereby leading to a decrease in the diffusion coefficient. In general, TAMS models have a significantly reduced diffusion coefficient than ZAMS and midMS models.

\subsection{Mass dependencies}\label{mass}
\begin{figure}
\includegraphics[width=0.45\textwidth]{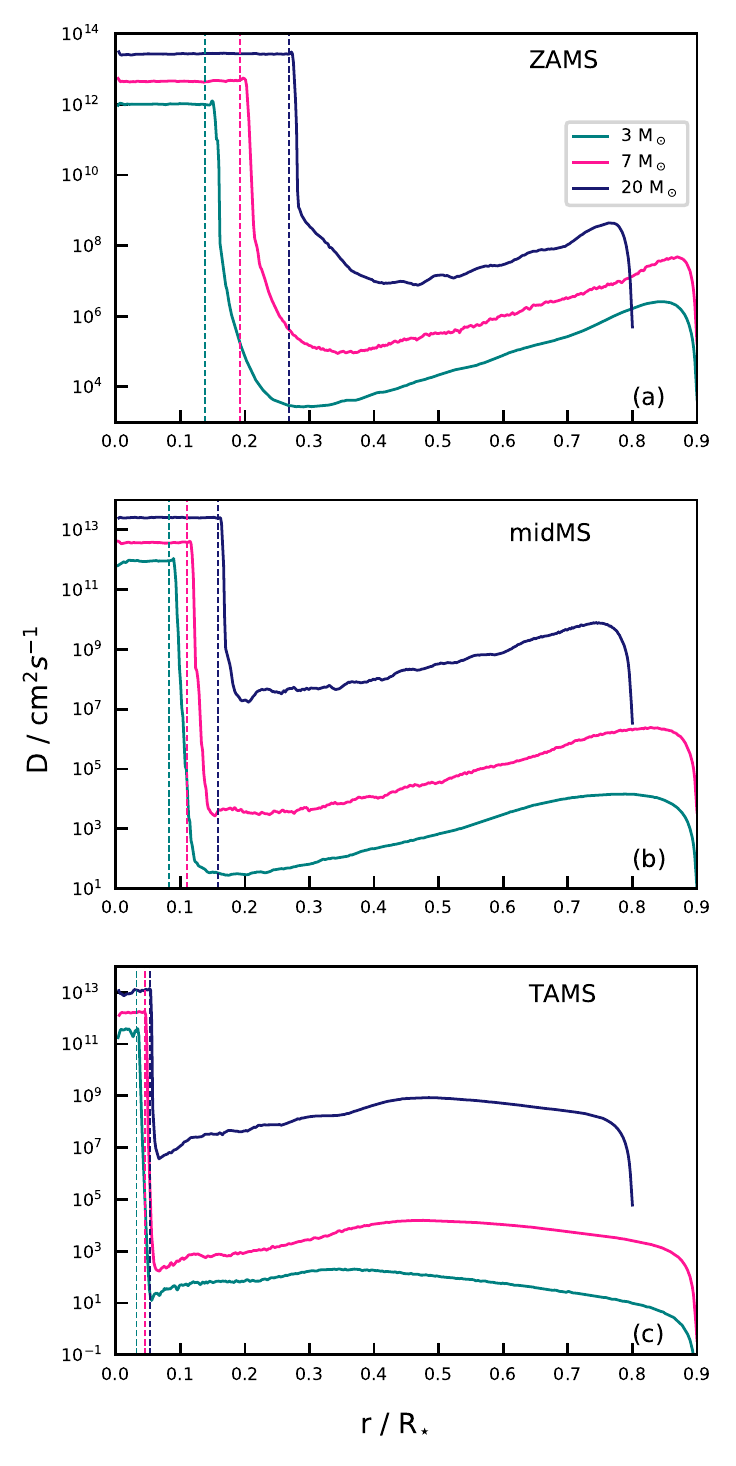}
\caption{The diffusion coefficient as a function of fractional radius for 3 M$_{\odot}$ , 7M$_{\odot}$ and 20 M$_{\odot}$ at (a) Zero Age Main sequence (b) mid - Main sequence and (c) Terminal Age Main Sequence. The vertical dashed line represents the convective-radiative interface for each models.}
\label{age_all}
\end{figure}

Fig. \ref{age_all} shows the diffusion profiles as a function of radius for 3 M$_{\odot}$ (green), 7M$_{\odot}$ (pink) and 20 M$_{\odot}$ (blue) at different ages. 
As evident from the figure, the mixing is stronger in massive stars, 
and this can be accounted for by the stronger convective driving in massive stars. Another factor contributing to the increase in the diffusion coefficient with mass is the Brunt-V\"{a}is\"{a}l\"{a} frequency. As noted from Fig. \ref{bvf_r}, the Brunt-V\"{a}is\"{a}l\"{a} frequency decreases with increasing mass. This leads to lower damping of waves in stars with higher mass and hence results in stronger mixing than lower mass stars.

\subsection{Parameterization of the Diffusion Coefficient}
\label{section:para}
\citetalias{Rogers2017} determined the diffusion profile in the radiation zone to be set by the square of the wave amplitude given by
\begin{align} \label{eq:1}
D =Av_{wave} ^{2} (\omega, l,r) ,
\end{align}
where the coefficient $A$ was speculated to depend on the diffusivities, rotation and dimensionality and was found to be $\sim$ $1$s by \citetalias{Rogers2017}.

To test this theory against the new data, we calculated the wave amplitudes using the linear theory given by \citet{rathish2019} where they solved the anelastic, linearised hydrodynamic equations within the WKB approximation without considering the rotational, thermal and viscous effects. The equation is given by 
\begin{align} \label{eq:2}
 v_{wave}(\omega,l,r) = v_{0}(\omega,l,r)\left(\frac{\rho}{\rho_{0}}\right)^{\frac{1}{2}}\left(\frac{r}{r_{0}}\right)^{-1} \\
 \left(\frac{N^{2}-\omega^{2}}{N_{0}^{2}-\omega^{2}}\right)^{-\frac{1}{4}}e^{-\frac{\mathcal{T}}{2}}, \nonumber
\end{align} 
where $\rho_0$, $r_0$ and $N_0$ are the density, radius and the Brunt-V\"{a}is\"{a}l\"{a} Frequency at the initial reference point. $v_{0}$($\omega$,l,r) is the initial wave amplitude for  a given frequency and wavenumber. The radial velocity is related to the tangential velocity by the following relation,
\begin{equation}
    \frac{v_{\theta}}{v_{r}} =\left(\frac{N^2}{\omega^2} -1 \right)^{\frac{1}{2}},
\end{equation}
The above can be approximated to $\frac{N}{\omega}$ for frequencies much smaller than the Brunt-V\"{a}is\"{a}l\"{a} Frequency. We therefore assume the wave amplitude to be equal to the root mean squared velocity at the reference point averaged over the time window considered (see Section \ref{2d}) multiplied by a factor of $\frac{\omega}{N}$ to obtain the theoretical radial diffusion profiles. 
The term $\exp({-\frac{\mathcal{T}}{2}})$ in the above equation is the attenuation factor multiplied to the wave amplitude as the radiative damping is considered within the quasi adiabatic limit \citep{zan97}. The damping coefficient $\mathcal{T}$ according to \citet{Kumar} is given as:
\begin{equation} \label{eq:3}
\hspace{-0.0cm}\mathcal{T} (\omega, l,r) =  \int_{r_{0}}^{r} \frac{16\sigma T^{3}}{3\rho^{2}\kappa c_{p}}\left(\frac{(l(l+1))^{\frac{3}{2}}N^{3}}{r^{3}\omega^{4}}\right)\left(1-\frac{\omega^{2}}{N^{2}}\right)dr.
\end{equation}
where $\sigma$, $T$, $c_p$, $\kappa$ and $l$ are the Stefan-Boltzmann constant, temperature, specific heat capacity at constant pressure, opacity and wavenumber. 
\citetalias{Rogers2017} studied a 3 M$_{\odot}$ midMS model considering up to 70 $\%$ 
of the total stellar radius without taking into account the effect of the Brunt-V\"{a}is\"{a}l\"{a} Frequency on the wave amplitude. We extend their analysis to stellar models presented in this work
by considering a modified relation to calculate the wave amplitude (Eqn. \ref{eq:2}) based on the linear theory given by \cite{rathish2019}.
\subsection{Determining the dominant frequencies contributing to the mixing profiles}\label{dom_freq}
\begin{figure}
\includegraphics[width=\columnwidth]{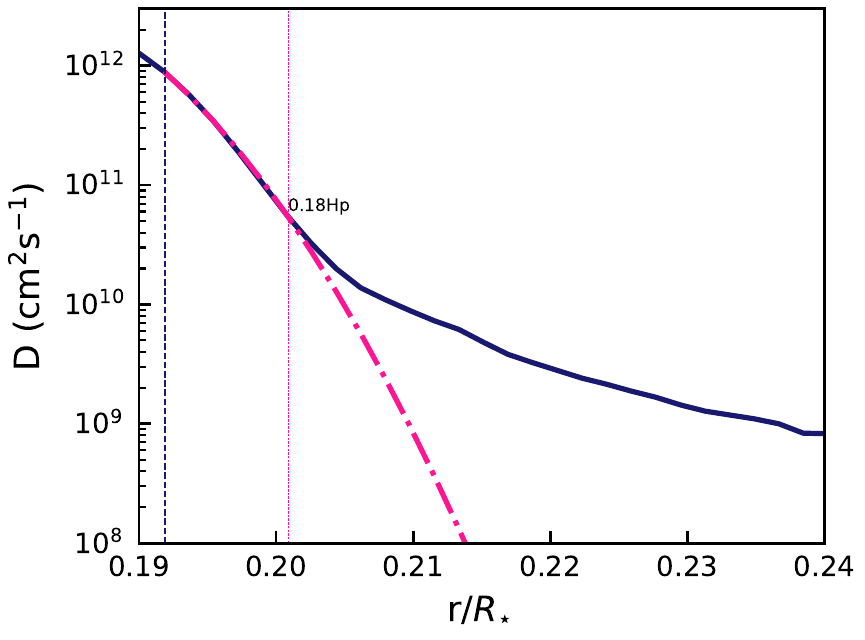}
\caption{Mixing profile near the convective radiative interface from the simulation (blue) and the Gaussian fit (pink line) with the overshooting depth (0.18H$_{p}$) indicated by the pink vertical dashed line for a 7 M$_{\odot}$ ZAMS model. The convective radiative interface is given by the vertical blue dashed line. }
\label{OD}
\end{figure}

In order to incorporate this mixing into 1D stellar evolution codes, we need to determine the dominant waves contributing to the mixing profiles, we launched waves from a radius outside the convective radiative interface and calculated the theoretical diffusion coefficient using Eqns. \ref{eq:1}-\ref{eq:3} for a set of frequencies ($1-40\mu $Hz) and wavenumbers (1-10) for each stellar model. 
The initial reference points where the waves are launched are highly influenced by the overshooting motions. We therefore initially determined the overshooting depth for each model  by fitting the following Gaussian function to the mixing profile near the convective radiative interface,
\begin{equation}
D = D_0\exp\left[\frac{(x-cr+f_{0}H_{p})^{2}}{2(f_{1}H_{p})^{2}}\right]
\label{gaus_OD}
\end{equation}
 where D$_0$ is the amplitude, {\it cr} is the radius of the convection zone, H$_{p}$ is the pressure scale height, $f_0$ and $f_1$ are the overshooting parameters. We measured the overshooting depth as the radius up to which we could get a good fit of the Gaussian profile. Fig. \ref{OD} shows the Gaussian fit along with the mixing profile near the convective radiative interface for a 7M$_{\odot}$ ZAMS model as an example. Here, we obtained the mixing profile (blue) near the convective radiative interface from the simulation 
 and then fitted this profile with the Gaussian function (pink) given by Eqn. \ref{gaus_OD}. The overshooting depth determined is indicated by the pink vertical dashed line. We repeated this procedure for all the models and the overshooting depth for the different models found are given in Table \ref{table_2}. We then launched the waves at a radius equal to 1.3 \footnote{ We considered various initial reference points beyond the overshoot depth to launch the waves and found that at this value, the theoretical profiles showed a reasonable agreement with that of the simulation profiles for all the models studied.} times the overshooting depth in all the models studied. More details on the overshooting profiles along with the comparison of the overshooting depth determined from the theory and simulation will be discussed in an upcoming paper.
\begin{figure*}
\includegraphics[width=1\textwidth]{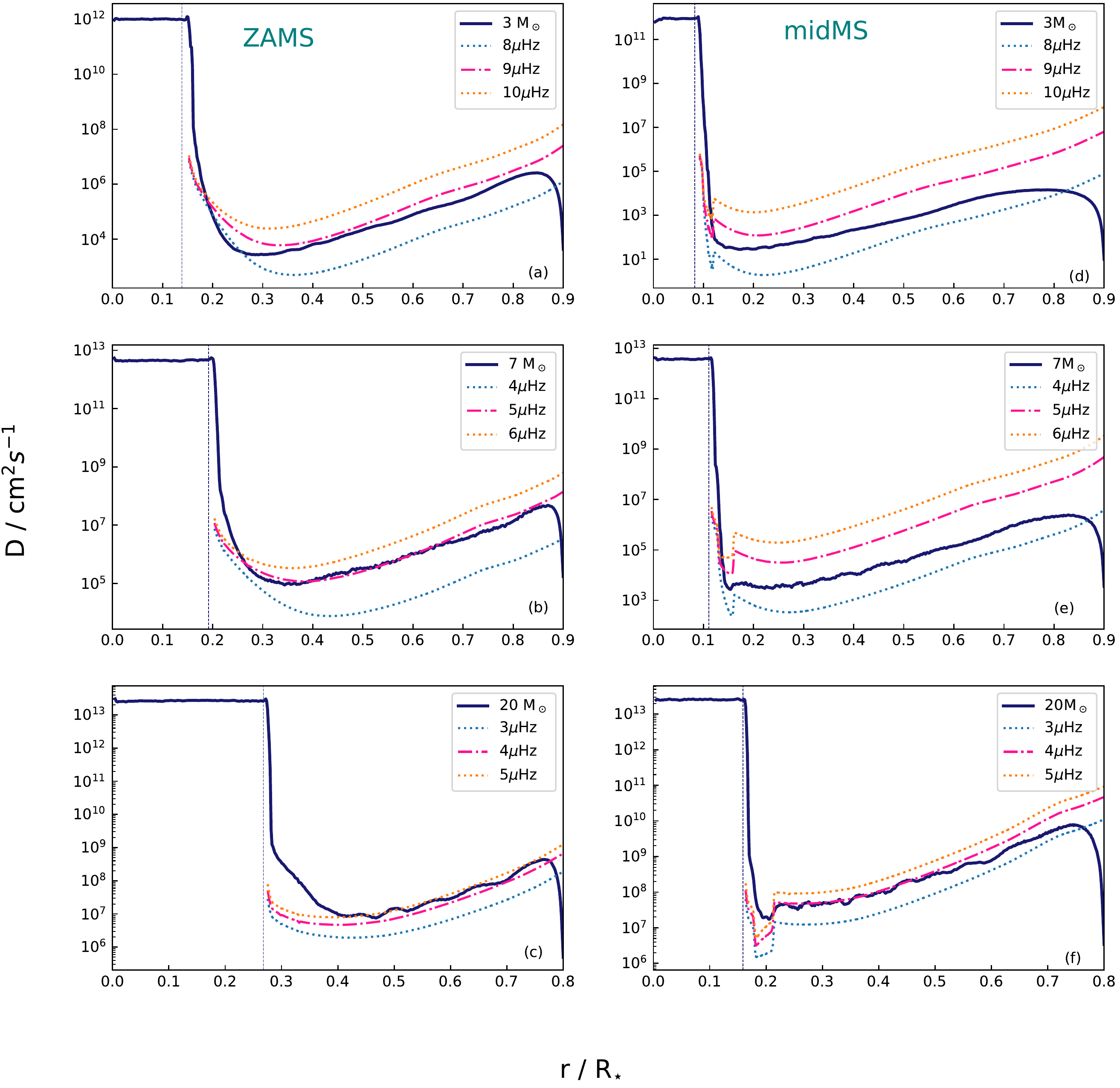}
\caption{Diffusion coefficient as a function of fractional radius for 3, 7 and 20 M$_{\odot}$ at ZAMS (left panel) and midMS models (right panel) from the simulation (blue line) and from the linear theory with waves launched at 1.3H$_{p}$ for  the dominant frequencies (pink line). The dotted profiles shows the theoretical profiles at different frequencies.}
\label{LT}
\end{figure*}

\begin{figure}
\includegraphics[width=0.45\textwidth]{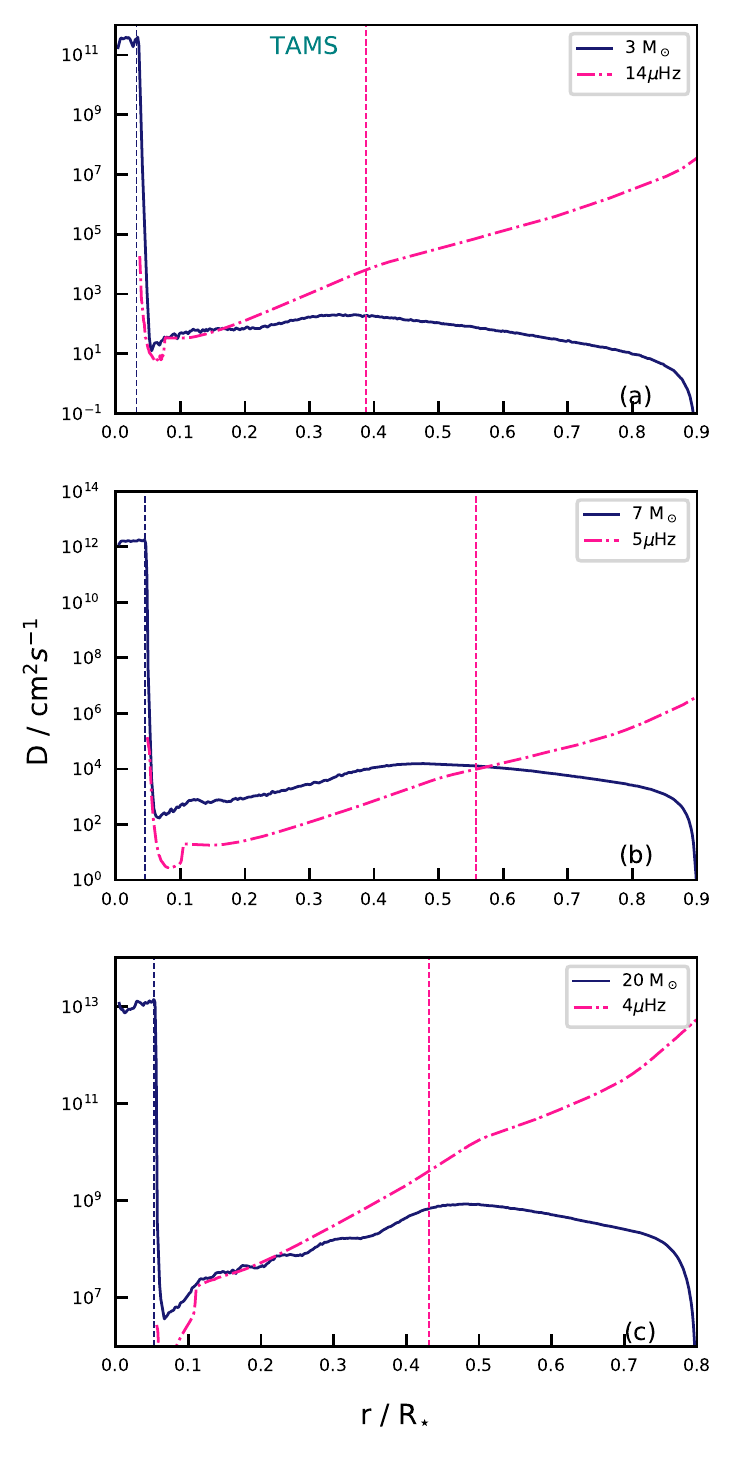}
\caption{Diffusion coefficient as a function of fractional radius for 3, 7 and 20 M$_{\odot}$ at TAMS from the simulation (blue line) and from the linear theory (pink line). 
The pink vertical dashed line indicates the turning point corresponding to the frequencies shown.}
\label{LT_TAMS}
\end{figure}


In order to implement a prescription for wave mixing into a 1D code, it would be helpful to narrow down the range of frequencies and wavenumbers contributing to the mixing profile. We therefore compared the theoretical profiles calculated using Eqns. \ref{eq:1}-\ref{eq:3} at different frequencies and wavenumbers with our simulation diffusion profiles  for all the models 
to determine 
if a simpler prescription, with limited waves, could be found. 
We observed that the waves with higher wavenumbers provide a smaller contribution towards the diffusion coefficient since they experience higher thermal damping as expected from Eqn. \ref{eq:3} and therefore, we  discard their contributions and consider only $l=1$ waves in our further calculations. 
We show the theoretical diffusion profiles corresponding to the dominant frequencies that provide the best fit to the profiles from the simulations in Fig. \ref{LT}. The figure also shows the theoretical profiles at different frequencies as an example (dotted profiles in Fig. \ref{LT}). In general, the mixing profiles obtained from the linear theory corresponding to the dominant frequencies are in good agreement with the simulations diffusion profiles for all ZAMS (Fig. \ref{LT} (a), (b), (c)) and midMS (Fig. \ref{LT} (d), (e), (f)) models. 


We found the dominant frequencies for the different models studied to be in the range of $4 - 9$ $\mu$Hz at wavenumber $l = 1$; with the lower frequencies being dominant for massive stars. This can be attributed to the thermal damping given by Eqn. \ref{eq:3}. Even though the lower frequencies are generated with larger amplitude near the convective-radiative interface, these waves are highly affected by thermal damping. As noted from Fig. \ref{bvf_r}, the average Brunt-V\"{a}is\"{a}l\"{a} frequency decreases with mass, causing 
the effect of damping to be less dominant in massive stars and thereby resulting in lower dominant frequencies.
We also noted that the waves experience extra damping near the convective radiative interface in all the midMS models because of the Brunt-V\"{a}is\"{a}l\"{a} frequency peak as noted from Fig. \ref{bvf_r}.
The theoretical profiles do not match the numerical simulations well at the surface, nor interface.  At the surface, this is because the radial velocity is forced to zero in our hydrodynamic simulations, causing the amplitude of our numerical profile to drop near the surface which is not necessarily physical. At the interface, convective overshooting means the region is not totally dominated by waves \citep{R2006}. In our models, the deviation of the theoretical diffusion profile from that of the simulation near the interface, particularly in the case of ZAMS can be explained by these overshooting motions, as they influence the determination of the initial reference point at which the waves are to be launched.

\begin{table}[ht]
  \centering
\begin{adjustbox}{width=0.4\textwidth}
\begin{tabular}{cc} 
 \hline
 Model & Overshooting Depth (Units of H$_{p}$)   \\ [0.5ex] 
 \hline
  3M$_{\odot}$ ZAMS&  0.22 \\
3M$_{\odot}$ midMS & 0.20 \\
3M$_{\odot}$ TAMS &0.17 \\
7M$_{\odot}$ ZAMS & 0.18 \\
7M$_{\odot}$ midMS & 0.105 \\
7M$_{\odot}$ TAMS &0.12 \\
20M$_{\odot}$ ZAMS & 0.14 \\

20M$_{\odot}$ midMS &0.11 \\
20M$_{\odot}$ TAMS &0.22 \\

 \hline
\end{tabular}
\end{adjustbox}
\caption{The overshooting depths in the units of pressure scale height (H$_{p})$ for the different models.}
\label{table_2}
\end{table}


We followed the same procedure to calculate the theoretical diffusion profiles for the TAMS models. We found that none of the theoretical profiles could explain the profile from the numerical simulation accurately. Fig. \ref{LT_TAMS}  shows  the best match that we could get from our analysis for the different TAMS models (pink dot-dashed profiles). 
The pink vertical dashed line represents the turning point for the frequencies shown.
We can clearly see that the profiles start to diverge as they approach the turning point beyond which the waves become evanescent.
As  stated previously, the waves lose their wave like behaviour at a lower fraction of the total stellar radius  because of the density term in Eqn. \ref{red_LT} being dominant for a major fraction of the radius in older stars. Therefore the WKB approximation applied to the equation for wave propagation (Eqn. \ref{red_LT}) is not valid in the TAMS models and hence the theoretical description outlined in \citetalias{Rogers2017} and Section \ref{section:para} is not valid in older stars. 
More analysis needs to be carried out in the future for a better understanding of the diffusion profiles of the TAMS models. However, at this stage it appears that a substantially lower diffusion coefficient, which is virtually flat throughout the radiation zone is a decent approximation of wave mixing at late stages of stellar evolution.

\begin{figure*}
\includegraphics[width=1\textwidth]{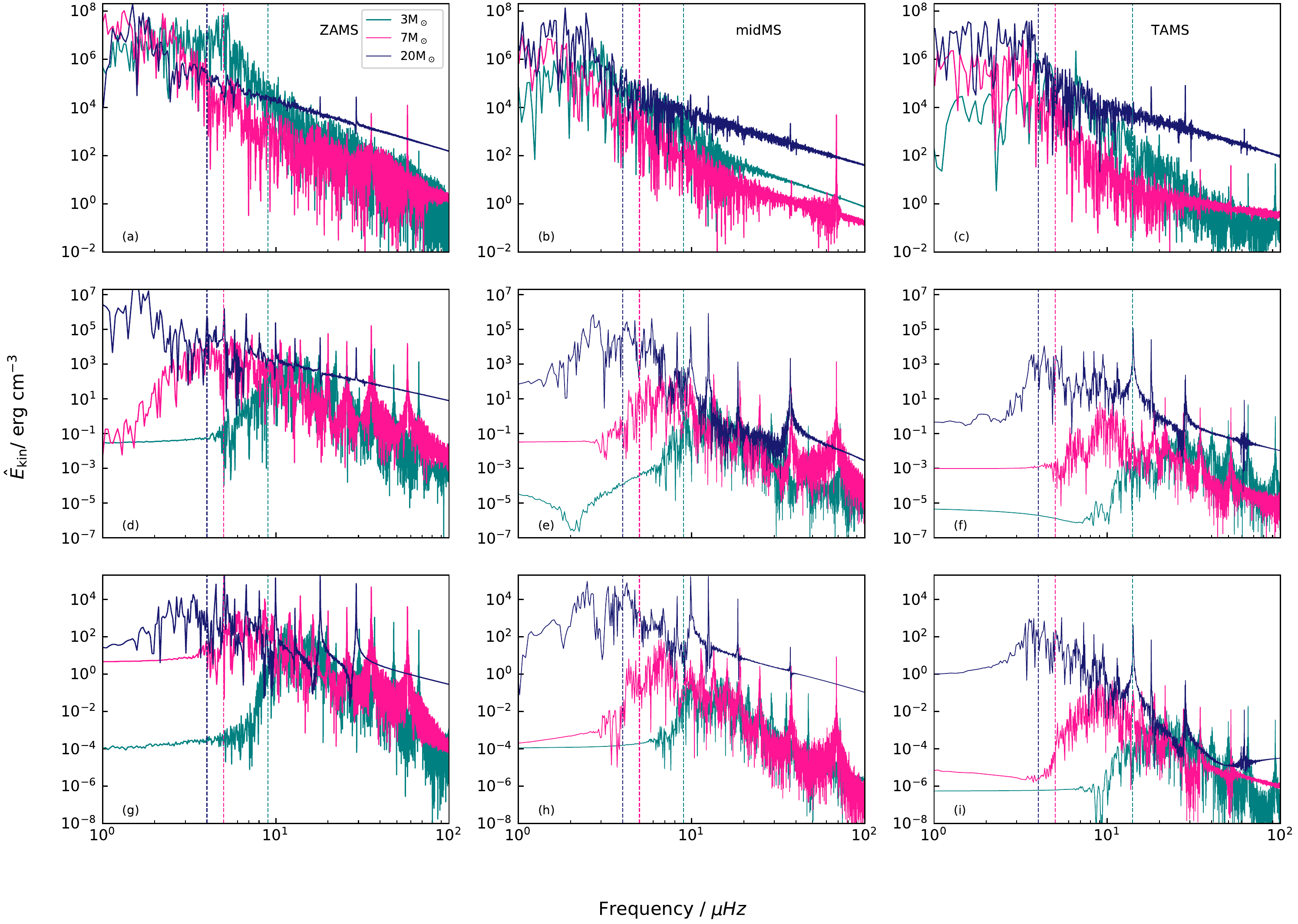}
    \caption{The frequency distribution of the kinetic energy density for 3 M$_{\odot}$ (green), 7 M$_{\odot}$ (pink) and 20 M$_{\odot}$ (blue) at ZAMS (left column), midMS (middle column) and TAMS (right column) in the radiation zone at 1.3 times the overshoot depth (top row), $0.3R_{\star}$ (middle row)  $0.6R_{\star}$ (bottom row). The vertical dashed lines indicate the dominant frequencies shown in Fig. \ref{LT} and Fig. \ref{LT_TAMS}.}
\label{KE}
\end{figure*}

We also computed the kinetic energy spectrum, $\hat{E}_{\mathrm{kin}}$,
\begin{equation}
    \hat{E}_{\mathrm{kin}}=\frac{1}{2} \bar{\rho}(\hat{v_r}^{2}+\hat{v_{\theta}}^{2})
\end{equation}
where, $\hat{v_{r}}$ and $\hat{v_{\theta}}$ are the Fourier transforms of radial  and tangential velocities. Fig. \ref{KE} shows the frequency distribution of the kinetic energy for $l=1$ at three different radii where the top panel (a, b, c) presents the spectra at 1.3 times the overshoot depth (radius at which the waves are launched),  the middle panel (d, e, f) at  $0.3R_{\star}$  and the bottom panel (g, h, i) at $0.6R_{\star}$ for all the models studied. The vertical dashed lines indicate the dominant frequencies shown in Fig. \ref{LT} and Fig. \ref{LT_TAMS}. We note that the dominant frequencies lie close to where the kinetic energy density peaks, thereby justifying our choice of frequencies. This is more evident from the middle and bottom panels of Fig. \ref{KE}. We are not concerned with the TAMS models as we have already shown that the linear theory does not explain their mixing profiles accurately. 
\vspace{0.5cm}
\subsection{Determination of the parameter A}\label{const_A}
\begin{figure*}
\includegraphics[width=1\textwidth]{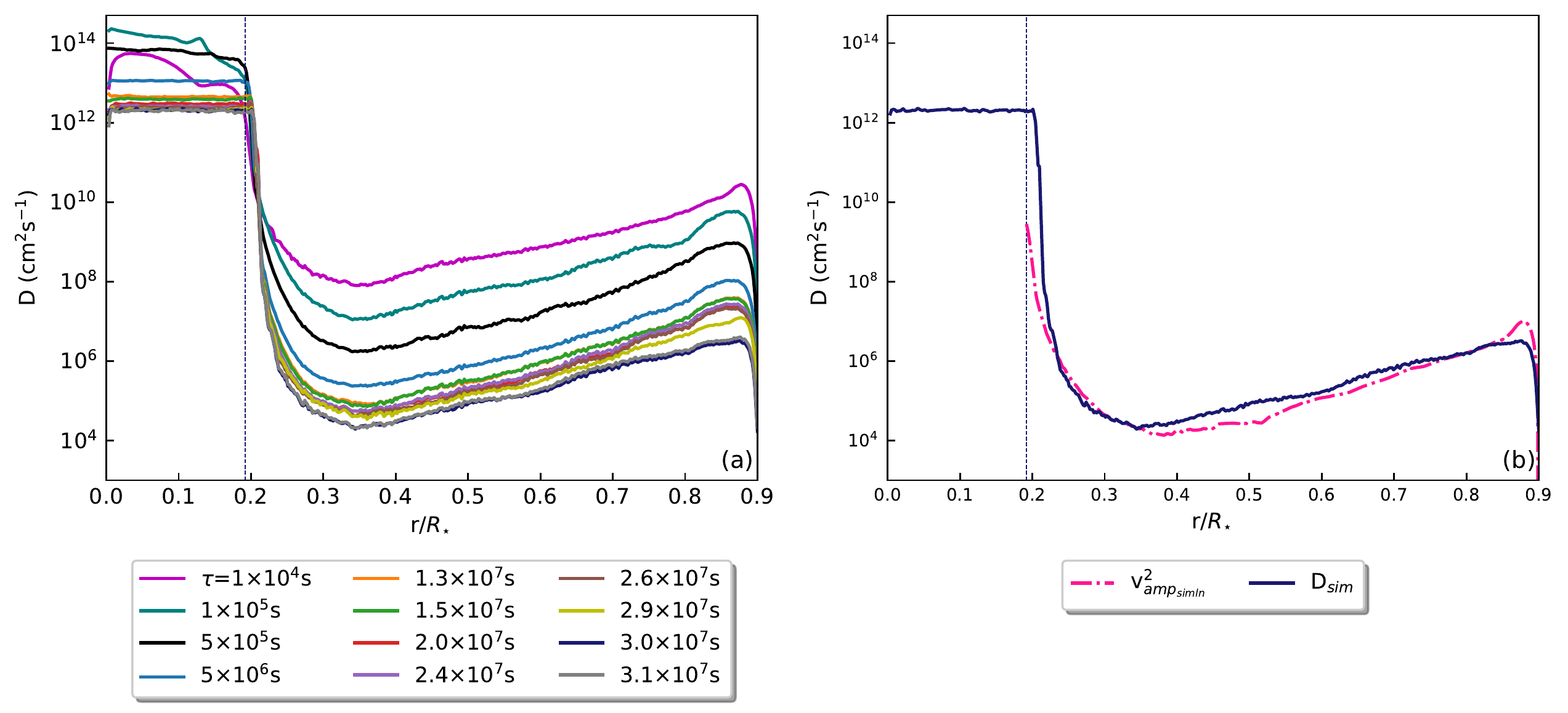}
\caption{(a) Radial diffusion profile for the 7 M$_{\odot}$ ZAMS model at different time differences ($\tau$). (b) The square of the wave amplitude (summed over frequencies ($1-40\mu $Hz) and wavenumbers (1-20)) from the simulation (pink) along with diffusion profile (blue) at $\tau=3 \times 10^7$s.}
\label{amp_sim_70}
\end{figure*}

We conducted a number of tests to determine the value of the parameter ’A’ from our simulations. We found that the simulation should be run for longer convective turnover times for the correct estimate of this parameter. Regardless of how long the simulations were run, the mixing profiles maintained the same trends discussed in Section \ref{age} and \ref{mass}. But, we noted that the amplitudes of the profiles decrease further and converge at higher time differences. This is shown in Fig. \ref{amp_sim_70}(a) for the 7 M$_{\odot}$ ZAMS model run for 87 convective turnover times. We see that the amplitude of the mixing profile converges at $\sim$ $\tau = 3\times 10^{7}$s. We then compared this profile to the square of wave amplitude (here the amplitude is summed over a range of frequencies ($1-40\mu $Hz) and wavenumbers (1-20)\footnote{The wave amplitudes were negligible at higher wavenumbers, hence we neglected their contributions in our calculation.}) obtained from the simulation (pink in Fig. \ref{amp_sim_70} (b)). As evident from the figure, both the profiles matches to a good extend, giving us the value of the parameter 'A' $\sim$ 1. We expect all the models to show similar behaviour once the simulation is run for longer convective turnover times. Our finding is in agreement with that of \citetalias{Rogers2017}. 
A better estimate of this parameter can only be obtained by the comparison of our theoretical prescription with that of the observations.
\section{Conclusion}\label{conclsn}
We studied the mixing by IGWs in stars of different masses and ages by introducing tracer particles into our 2D hydrodynamic simulations. Initially, we carried out a convergence test for all our models to check if they agreed with the studies of \citetalias{Rogers2017}. We found that all our models attained numerical convergence with respect to the time, particle and radial resolution. 
We then focused our attention on the diffusion profiles as a function of mass and age. 

One of our main conclusions is that the mixing profile varies significantly across age with the  mixing stronger in younger stars. The different trend observed in the mixing profiles of TAMS models can be accounted for by the turning point which is located at a lower fraction of the total stellar radius and the Brunt-V\"{a}is\"{a}l\"{a} frequency spike in these older stars. We also found that mixing by IGWs is stronger for stars of higher masses. The stronger convective driving in massive stars along with the lower average Brunt-V\"{a}is\"{a}l\"{a} frequency result in higher amplitude waves within the radiative zone and therefore, more mixing.

We then tested our simulation results against the prescription given by \citetalias{Rogers2017}
for the inclusion of these mixing profiles in one dimensional stellar evolution models.
We found that their theory agrees well with the simulation for all ZAMS and midMS models
with the dominant waves contributing to these mixing profiles in the range of $4 - 9 \mu \mathrm{Hz}$ for wavenumber $l=1$. We also found that the dominant frequency decreases with increasing mass, but remains approximately the same across different ages for a given mass as can be noted from Fig. \ref{LT}. When run for longer time, we find that the amplitude of the mixing profile converges such that A $\sim$ 1. It is still unclear what physically sets this parameter. The linear theory however failed to explain the mixing profiles of TAMS models because the waves cannot be described by the WKB approximation. Studies attempting to find a prescription for the TAMS models will be forthcoming. 

Future work will focus on determining the  effect of rotation on the mixing by IGWs in the radiation zone by extending our analysis to models of different rotation rates. We aim to find out how rotation influences the turning point in older stars. 
\section{Acknowledgements}\label{ack}
We acknowledge support from STFC grant ST/L005549/1 and
NASA grant NNX17AB92G. Computing was carried out on Pleiades at NASA Ames, Rocket High Performance Computing service at Newcastle University and DiRAC Data Intensive service at Leicester, operated by the University of Leicester IT Services, which forms part of the STFC DiRAC HPC Facility (\url{www.dirac.ac.uk}), funded by BEIS capital funding via STFC capital grants ST/K000373/1 and ST/R002363/1 and STFC DiRAC Operations grant ST/R001014/1. PVFE was supported by the U.S. Department of Energy through the Los Alamos National Laboratory (LANL). LANL is operated by Triad National Security, LLC, for the National Nuclear Security Administration of the U.S. Department of Energy (Contract No. 89233218CNA000001). This paper was assigned a document release number LA-UR-22-27682.
\appendix 
\section{MESA inlist files}\label{app}
The MESA inlists used to generate the models in this work are given below.  To keep comparisons consistent, we just changed the mass and the value of Xc (xa\textunderscore central\textunderscore lower\textunderscore limit(1)) in the inlists without changing any other parameters.

The macro and micro physics details are implemented using inlist\textunderscore project:
\vspace{0.2cm}
\begin{verbatim}
&star_job
! mesa_dir = '$MESA_DIR'
create_pre_main_sequence_model = .false.
show_log_description_at_start = .false. 
show_net_species_info = .false.
show_net_reactions_info = .false.      
show_eqns_and_vars_names = .false.      
!history_columns_file = 'hist.list' 
!profile_columns_file = 'prof.list' 
report_retries = .false.
report_backups = .false.
initial_zfracs = 6         
change_v_flag = .true.
new_v_flag = .true.
change_lnPgas_flag = .true.
new_lnPgas_flag = .true.
change_net = .true. 
! switch nuclear reaction network
new_net_name = 'o18_and_ne22.net'   
auto_extend_net = .true.
kappa_file_prefix = 'a09' 
kappa_lowT_prefix = 'lowT_fa05_a09p' 
kappa_blend_logT_upper_bdy = 4.10 
kappa_blend_logT_lower_bdy = 4.00 
kappa_type2_logT_lower_bdy = 3.8 
kappa_CO_prefix = 'a09_co' 
/ !end of star_job   

&controls
stop_near_zams = .true.
! Initial values
initial_mass = 3.00
initial_z    = 2d-2  
!initial_y = 0.58
!zams_filename = 'zams_2m2_y58.data'
 
! controls for output         
terminal_interval = 25 
write_header_frequency = 5
terminal_show_age_in_years = .true. 
photo_interval = 200
photo_digits = 4 
!log_directory = 'LOGS' 
history_interval = 1       
write_profiles_flag = .false. 
profile_interval = 100000 
priority_profile_interval = 10000
profile_data_suffix = '.prof'
star_history_dbl_format = '(1pes22.6e3, 1x)'
star_history_int_format = '(i22, 1x)'
star_history_txt_format = '(a32, 1x)'
!max_model_number = 1
xa_central_lower_limit_species(1) = 'h1' 
xa_central_lower_limit(1) = 0.01d0 
! 0.69d0, 0.35d0, 0.01
when_to_stop_rtol = 1d-4
when_to_stop_atol = 1d-4
max_num_profile_models = -1  
               
write_pulse_info_with_profile = .false. 
!pulse_info_format = 'FGONG'            
add_atmosphere_to_pulse_info = .true.
add_center_point_to_pulse_info = .false. 
keep_surface_point_for_pulse_info = .true. 
         
! mixing parameters                 
 use_Ledoux_criterion = .true.
!use_Henyey_MLT = .true.
 MLT_option = 'Henyey' 
 mixing_length_alpha = 1.8
 ! alpha_semiconvection = 0 ! 0.01 
 num_cells_for_smooth_gradL_composition_term = 3
  radiation_turbulence_coeff = 0 
      
! apply smoothing to abundances in newly
nonconvective regions
smooth_convective_bdy = .true.
max_delta_limit_for_smooth = 0.1 
mixing_D_limit_for_log = 1d-10

! Overshooting
overshoot_f_above_burn_h_core = 0.020
overshoot_f0_above_burn_h_core = 0.001
D_mix_ov_limit = 1d0
   
! atmosphere boundary conditions

which_atm_option = 'photosphere_tables'                       
atm_switch_to_grey_as_backup = .true.            
! parameters for Paczynski_grey
Paczynski_atm_R_surf_errtol = 1d-4
create_atm_max_step_size = 0.01 

! mass gain or loss

mdot_omega_power = 0.43 
max_rotational_mdot_boost = 1d4 
hot_wind_scheme = 'Dutch'
Dutch_scaling_factor = 1d0 
Dutch_wind_lowT_scheme = 'van Loon'
mass_change_full_on_dt = 3.15d8  
mass_change_full_off_dt = 3.15d7 

! mesh adjustment
! Initial setting for mesh, but for the 3rd run, 
I use inlist_zoom.

      
use_other_mesh_functions = .true.

T_function2_weight = 100
T_function2_param = 18d4      

xtra_coef_above_xtrans = 0.5
xtra_coef_below_xtrans = 0.5
xtra_dist_above_xtrans = 0.4
xtra_dist_below_xtrans = 0.4
      
mesh_dlog_pp_dlogP_extra = 0.20
mesh_dlog_cno_dlogP_extra = 0.20

mesh_dlog_3alf_dlogP_extra = 0.20
mesh_dlog_burn_c_dlogP_extra = 0.20
mesh_dlog_burn_n_dlogP_extra = 0.20
mesh_dlog_burn_o_dlogP_extra = 0.20

! timestep controls
dX_nuc_drop_limit = 5d-4
dX_nuc_drop_hard_limit = 1d-3

!nuclear reaction controls      
default_net_name = 'cno_extras.net'

! element diffusion 
do_element_diffusion = .false. 
diffusion_dt_limit = 3.15d7 
               
! opacity controls                     
use_Type2_opacities = .false.
Zbase = 2d-2

! asteroseismology controls

! for calculations of delta_nu and nu_max
get_delta_nu_from_scaled_solar = .true.
nu_max_sun = 3100d0
delta_nu_sun = 135d0
Teff_sun = 5777d0
                  
! eps_grav            
use_lnS_for_eps_grav = .true.            
report_ierr = .false. 

report_why_dt_limits = .false.
         
! Brunt
calculate_Brunt_N2 = .true.
num_cells_for_smooth_brunt_B = 3
use_brunt_gradmuX_form = .false. 
/
\end{verbatim}
Mesh adjustments are given by inlist\textunderscore zoom:
\begin{verbatim}
    
&star_job

/


&controls

!max star age
max_age = 700d7

! mesh adjustment
mesh_delta_coeff = 0.5
max_allowed_nz = 2000000      
okay_to_remesh = .true.
          

! radius gradient
R_function_weight = 1!0
R_function_param = 1d-4
R_function2_weight = 1000!10
R_function2_param1 = 0.8!1000
R_function2_param2 = 0.7!0

xtra_coef_above_xtrans = 0.1
xtra_coef_below_xtrans = 0.1
xtra_dist_above_xtrans = 0.5
xtra_dist_below_xtrans = 0.5
mesh_logX_species(1) = 'h1' 
mesh_logX_min_for_extra(1) = -6
mesh_dlogX_dlogP_extra(1) = 0.1
mesh_dlogX_dlogP_full_on(1) = 2 
mesh_dlogX_dlogP_full_off(1) = 1 

mesh_logX_species(2) = 'he4' 
mesh_logX_min_for_extra(2) = -6      
mesh_dlogX_dlogP_extra(2) = 0.1
mesh_dlogX_dlogP_full_on(2) = 2 
mesh_dlogX_dlogP_full_off(2) = 1 
       
xtra_coef_czb_full_on = 1d0 
xtra_coef_czb_full_off = 1d0 

xtra_coef_a_l_nb_czb = 0.1 
xtra_coef_a_l_hb_czb = 0.1 
xtra_coef_a_l_heb_czb = 0.1 
xtra_coef_a_l_zb_czb = 0.1 

xtra_coef_b_l_nb_czb = 0.1 
xtra_coef_b_l_hb_czb = 0.1 
xtra_coef_b_l_heb_czb = 0.1 
xtra_coef_b_l_zb_czb = 0.1 

xtra_coef_a_u_nb_czb = 0.1 
xtra_coef_a_u_hb_czb = 0.1 
xtra_coef_a_u_heb_czb = 0.1 
xtra_coef_a_u_zb_czb = 0.1 

xtra_coef_b_u_nb_czb = 0.1 
xtra_coef_b_u_hb_czb = 0.1 
xtra_coef_b_u_heb_czb = 0.1 
xtra_coef_b_u_zb_czb = 0.1 

xtra_dist_a_l_nb_czb = 0.5 
xtra_dist_a_l_hb_czb = 0.5 
xtra_dist_a_l_heb_czb = 0.5 
xtra_dist_a_l_zb_czb = 0.5 

xtra_dist_b_l_nb_czb = 0.5 
xtra_dist_b_l_hb_czb = 0.5 
xtra_dist_b_l_heb_czb = 0.5 
xtra_dist_b_l_zb_czb = 0.5 

xtra_dist_a_u_nb_czb = 0.5 
xtra_dist_a_u_hb_czb = 0.5 
xtra_dist_a_u_heb_czb = 0.5 
xtra_dist_a_u_zb_czb = 0.5 

xtra_dist_b_u_nb_czb = 0.5 
xtra_dist_b_u_hb_czb = 0.5 
xtra_dist_b_u_heb_czb = 0.5 
xtra_dist_b_u_zb_czb = 0.5 


xtra_coef_os_full_on = 1d0 
xtra_coef_os_full_off = 1d0 

xtra_coef_os_above_nonburn = 0.1
xtra_coef_os_below_nonburn = 0.1
xtra_coef_os_above_burn_h = 0.1
xtra_coef_os_below_burn_h = 0.1   
xtra_coef_os_above_burn_he = 0.1
xtra_coef_os_below_burn_he = 0.1
xtra_coef_os_above_burn_z = 0.1
xtra_coef_os_below_burn_z = 0.1


xtra_dist_os_above_nonburn = 0.5
xtra_dist_os_below_nonburn = 0.5      
xtra_dist_os_above_burn_h = 0.5
xtra_dist_os_below_burn_h = 0.5      
xtra_dist_os_above_burn_he = 0.5
xtra_dist_os_below_burn_he = 0.5     
xtra_dist_os_above_burn_z = 0.5
xtra_dist_os_below_burn_z = 0.5


convective_bdy_weight = 0
convective_bdy_dq_limit = 1d-4 
convective_bdy_min_dt_yrs = 1d-3

okay_to_remesh = .true. 
       
remesh_dt_limit = -1 
remesh_log_L_nuc_burn_min = -50

/ ! end of controls namelist

\end{verbatim}

\bibliography{sample631}{}
\bibliographystyle{aasjournal}



\end{document}